\documentclass[runningheads]{llncs}
\usepackage{graphicx}
\usepackage{amsmath,amssymb} 
\usepackage{color}
\usepackage[width=122mm,left=12mm,paperwidth=146mm,height=193mm,top=12mm,paperheight=217mm]{geometry}
\usepackage{enumitem}
\usepackage{pifont}
\usepackage{booktabs}
\usepackage{multirow}

\newcommand{\figref}[1]{Figure~\ref{#1}}
\newcommand{\tabref}[1]{Table~\ref{#1}}

\newcommand{\secref}[1]{Section~\ref{#1}}
\newcommand{\comment}[1]{\textcolor{green}{}}

\newcommand\T{\rule{0pt}{2.6ex}}			
\newcommand\B{\rule[-1.2ex]{0pt}{0pt}} 	

\begin{document}
\pagestyle{headings}
\mainmatter
\def\ECCV18SubNumber{1047}  

\title{CeMNet: Self-supervised learning for \\ accurate continuous ego-motion estimation}

\titlerunning{CeMNET: Self-supervised learning for continuous ego-motion}


\author{Minhaeng Lee and Charless C. Fowlkes}
\institute{Dept. of Computer Science, University of California, Irvine}

\maketitle

\begin{abstract}
In this paper, we propose a novel self-supervised learning model for estimating
continuous ego-motion from video. Our model learns to estimate camera motion by
watching RGBD or RGB video streams and determining translational and rotation
velocities that correctly predict the appearance of future frames.  Our
approach differs from other recent work on self-supervised
structure-from-motion in its use of a continuous motion formulation and
representation of rigid motion fields rather than direct prediction of camera
parameters. To make estimation robust in dynamic environments with multiple
moving objects, we introduce a simple two-component segmentation process that
isolates the rigid background environment from dynamic scene elements. We
demonstrate state-of-the-art accuracy of the self-trained model on several
benchmark ego-motion datasets and highlight the ability of the model to provide
superior rotational accuracy and handling of non-rigid scene motions.

\end{abstract}

\section{Introduction}
Supervised machine learning techniques based on deep neural networks have shown
remarkable recent progress for image recognition and segmentation tasks.
However, progress in applying these powerful methods to geometric tasks such as
structure-from-motion has been somewhat slower due to a number of factors.  One
challenge is that standard layers defined in convolutional neural network (CNN)
architectures do not offer a natural way for researchers to incorporate
hard-won insights about the algebraic structure of geometric vision problems,
instead relying on general approximation properties of the network to
re-discover these facts from training examples.  This has resulted in some
development of new building blocks (layers) specialized for geometric
computations that can function inside standard gradient-based optimization
frameworks (see e.g., \cite{handa2016ECCV,Huang2017CVPR}) but
interfacing these to image data is still a challenge.  

A second difficulty is that optimizing convolutional neural networks (CNNs)
requires large amounts of training data with ground-truth labels. Such
ground-truth data is often not available for geometric problems (i.e., often
requires expensive special-purpose hardware rather than human annotations).
This challenge has driven recent effort to develop more realistic synthetic
datasets such as Flying Chairs and MPI-Sintel \cite{Butler2012ECCV} for flow and disparity
estimation, Virtual KITTI \cite{Gaidon2016CVPR} for object detection and
tracking, semantic segmentation, flow and depth estimation, 
and SUNCG \cite{song2016ssc} for indoor room
layout, depth and normal estimation.

In this paper, we overcome some of these difficulties by taking a
``self-supervised'' approach to learning to estimate camera motions
directly from video.  Self-supervision utilizes unlabeled image data by
constructing an encoder that transforms the image into an alternate
representation and a decoder that maps back to the original image.  This
approach has been widely for low-level synthesis problems such as
super-resolution~\cite{Dong2016TPAMI}, image colorization~\cite{zhang2016ECCV} 
and in-painting~\cite{pathak2016CVPR}
where the encoder is fixed (creating a downsampled, grayscale or occluded
version of the image) and the decoder is trained to reproduce the original
image. For estimation tasks such as human pose~\cite{tung2017NIPS},
depth~\cite{Vijayanarasimhan17Corr,Zhou2017cvpr}, and intrinsic image decomposition~\cite{janner2017NIPS}, the
structure of the decoder is typically specified by hand (e.g., synthesizing the
next video frame in a sequence based on estimated optical flow and previous
video frame) and the encoder is learned.  This framework is appealing for
geometric estimation problems since (a) it doesn't require human supervision to
generate target labels and hence can be trained on large, diverse data, and (b)
the predictive component of the model can incorporate user insights into the
problem structure.

Our basic model takes a pair of calibrated RGB or RGBD video frames as input,
estimates optical flow and depth, determines camera and object velocities, and
resynthesizes the corresponding motion fields. We show that the model can be
trained end-to-end with a self-supervised loss that enforces consistency of the
predicted motion fields with the input frames yields a system that provides
highly accurate estimates of camera ego-motion.  We measure the effectiveness
of our method using TUM~\cite{sturm2012iros} and Virtual
KITTI~\cite{Gaidon2016CVPR} dataset. 

Relative to other recent
papers~\cite{Tung2017Corr,Vijayanarasimhan17Corr,Zhou2017cvpr} that have also
investigated self-supervision for structure-from-motion, the novel
contributions of our work are:
\vspace{-0.1in}
\begin{itemize}[label=$\bullet$]
\item{We represent camera motion implicitly in terms of motion fields and depth
which are a better match for CNNs architectures that naturally operate in the
image domain (rather than camera parameter space). We demonstrate that this 
choice yields better predictive performance, even when trained in the fully
supervised setting}
\item{Unlike previous self-supervised techniques, our model uses a continuous
(linearized) approximation to camera
motion~\cite{Oliensis2004eccv,Jaegle2016icra} which suitable for video odometry
and allows efficient backpropagation while providing strong constraints for
learning from unsupervised data.}
\item{Our experimental results demonstrate state-of-the-art performance on
benchmark datasets which include non-rigid scene motion due to dynamic objects.
Our model improves on substantially on estimates of camera rotation, suggesting
this approach can serve well as a drop-in replacement for local estimation in
existing RGB(D) SLAM pipelines.}
\end{itemize}

\begin{figure}[t]
\begin{center}
\begin{tabular}{c}
\includegraphics[trim=0 450 0 50, clip, scale=0.23]{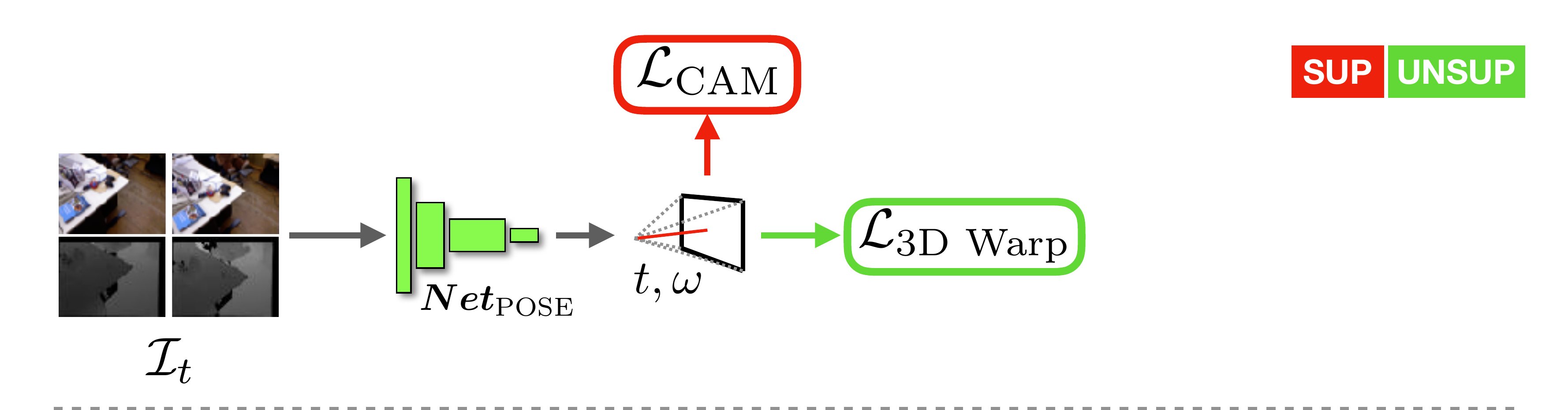}\\
\includegraphics[trim=0 0 0 40, clip, scale=0.23]{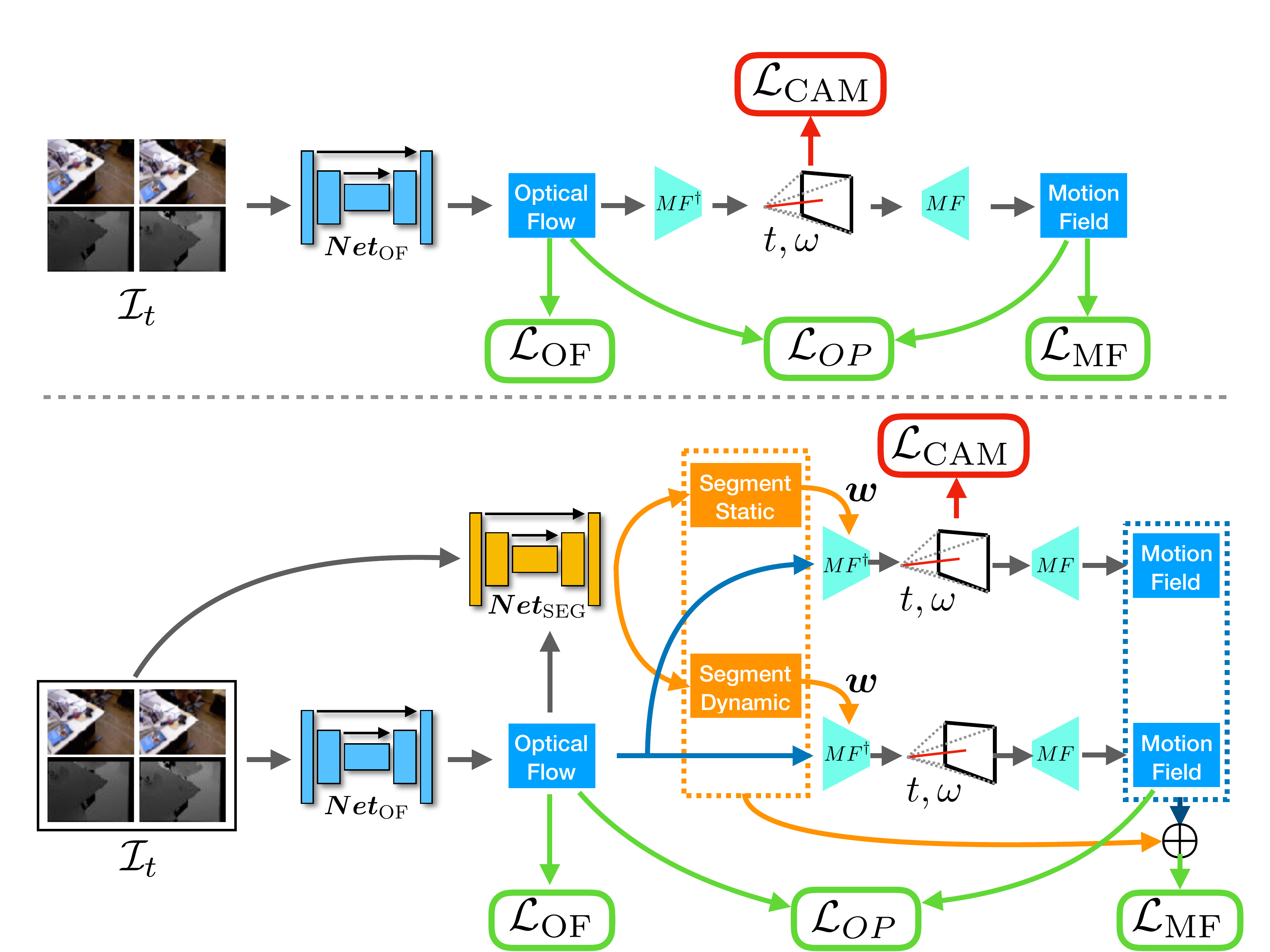}\vspace{-0.8em}
\end{tabular}
\caption{Overview of network architectures used in our experiments. The top
panel shows conventional (\emph{baseline}) approach that directly predicts 6DoF
camera motion.  The middle panel displays our proposed \emph{single layer
model} which predicts ego motion assuming a static (rigid) environment. Note
that our model supports both supervised (red) and unsupervised (green) losses
during training. The bottom panel shows a \emph{two layered model} variant that
segments a scene into static and dynamic components and only uses static
component for camera motion prediction. When input depth is not available, we
utilize an additional monocular depth estimation network to predict it (not
shown).}
\label{fig:overview}
\end{center}
\vspace{-0.1in}
\end{figure}

\section{Related Work}
Visual odometry is a classic and well studied problem in computer vision. Here
we mention a few recent works that are most closely related to our approach.

\noindent \textbf{Optical Flow, Depth and Odometry:} 
A number of recent papers have shown great success in estimation of optical
flow from video using learning-based
techniques~\cite{Dosovitskiy2015iccv,Ilg2016corr}.
Ren \emph{et al.}
introduced unsupervised learning for optical flow prediction~\cite{Ren2017aaai}
using photometric consistency.  Garg et al. utilize consistency between stereo
pairs to learn monocular depth estimation in a self-supervised manner
\cite{Garg2016eccv}.  Zhou et al.~\cite{Zhou2017cvpr} jointly trains estimators for
monocular depth and relative pose using an unsupervised loss.
SfM-Net~\cite{Vijayanarasimhan17Corr} takes a similar approach but explicitly
decomposes the input into multiple motion layers.  ~\cite{Ruihao2017arXiv} uses
stereo video for joint training of depth and camera motion (sometimes referred
to as scene flow) but tests on monocular sequences. Our approach differs from
these recent papers in using a continuous formulation appropriate for video.
Such a formulation was recently used by Jaegle \emph{et
al.}\cite{Jaegle2016icra} for robust monocular ego-motion estimation but using
classic (sparse) optical flow as input.

\noindent \textbf{SLAM:} While conventional simultaneous localization and
mapping (SLAM) methods estimate geometric information by extracting feature
points~\cite{kerl13iros,WhelanRSS15} or use all information in the given
images~\cite{Engel2014ECCV}, recently several learning-based methods have been
introduced. Tateno \emph{et al.}~\cite{tateno2017cvpr} propose a fusion SLAM
technique by utilizing CNN based depth map prediction and monocular SLAM.
Melekhov \emph{et al.} propose CNN based relative pose estimation using
end-to-end training with a spatial pyramid pooling
(SPP)~\cite{melekhov2017arxiv}.  Other recent works~\cite{Kim2016TOR,li2017rgb}
model static background to predict accurate camera pose even in dynamic
environment. Sun \emph{et al.} try to solve dynamic scene problem by adding
motion removal approach as a pre-processing to be integrated into RGBD
SLAM~\cite{SUN2017RAS}. Finally, the work of Wang \emph{et
al.}~\cite{Wang2017icra} train a recurrent CNN to capture longer-term
processing of sequences typically handled by bundle adjustment and loop
closure.

\section{Continuous Ego-motion Network}

~\figref{fig:overview} provides an overview of three different types of
architectures we consider in this paper.  We take as input a successive pair of
RGB images $\{I_t, I_{t+\delta}\}$ and corresponding depth images $\{d_t,
d_{t+\delta}\}$. When depth is not available, we assume it is predicted by a
monocular depth estimator (not shown).  The first network,
$\boldsymbol{Net}_{\text{POSE}}$, directly predicts 6 DoF camera motion by
attaching several fully connected layers at the end of several CNN layers. When
camera motion is known, this baseline can be trained with a supervised loss
${\mathcal L}_{\text{CAM}}$ or trained with a self-supervised image warping
loss ${\mathcal L}_{\text{3DWARP}}$ as done in several recent papers
~\cite{Zhou2017cvpr,Tung2017Corr,Vijayanarasimhan17Corr}.  

Instead of directly predicting camera motion, we advocate utilizing a
fully-convolutional encoder/decoder architecture with skip connections (e.g.,
~\cite{Ronneberger2015miccai,Shelhamer2017TPAMI,Dosovitskiy2015iccv,Ilg2016corr}) to first predict
optical flow (denoted $\boldsymbol{Net}_{\text{OF}}$). We then estimate
continuous ego-motion $(t,\omega)$ using weighted least-squares and
resynthesize the corresponding motion field $MF(t,\omega)$. These intermediate
representations can be learned using unsupervised losses (${\mathcal
L}_{\text{OF}}$,${\mathcal L}_{\text{MF}}$, ${\mathcal L}_{\text{OP}}$)
described below.  When additional moving objects are present in the scene,
we introduce an additional segmentation network, $\boldsymbol{Net}_{\text{SEG}}$,
which decomposes the optical flow into layers which are fit separately.

In the following sections we develop the continuous motion formulation,
interpret our model as projecting the predicted optical flow on to the subspace
of ego-motion flows, and discuss implementation of segmentation into layers.

\subsection{Estimating Continuous Ego-motion}
\label{sec:continuous_formulation}
Consider the 2D trajectory of a point in the image $\boldsymbol{x} = \{ x, y
\}$ as a function of its 3D position $\boldsymbol{X} = \{ X, Y, Z \}$
and motion relative to the camera. We write
\begin{eqnarray}
\boldsymbol{x}(t) &=& \{ x(t), y(t) \} = \bigg\{ \frac{fX(t)}{Z(t)}, \frac{fY(t)}{Z(t)} \bigg\} \nonumber,
\end{eqnarray} where $f$ is the camera focal length.
To compute the projected velocity in the image $\boldsymbol{v}(\boldsymbol{x})
= (v_x, v_y)^\top \in \mathbb{R}^2$ as a
function of the 3D velocity $\boldsymbol{V}(\boldsymbol{X})$ we take partial
derivatives.  For example, the $x$ component of the velocity is:
\begin{eqnarray}
\frac{\partial x(t)}{\partial t} &=& \frac{f}{Z(t)}\frac{\partial X(t)}{\partial t} + fX(t)\cdot \frac{\partial}{\partial t}\frac{1}{Z(t)}\nonumber\\
&=&\frac{f}{Z(t)}V_x - fX(t) \cdot \frac{1}{Z^2(t)}\cdot\frac{\partial Z(t)}{\partial t}\nonumber\\
&=& \frac{1}{Z(t)}
\begin{bmatrix}
f & 0 & -x(t)
\end{bmatrix}
\begin{bmatrix}
V_x \\
0 \\
V_z
\end{bmatrix} \nonumber
\end{eqnarray}
Dropping $t$ for notational simplicity, we can thus write the image velocity as:
\begin{eqnarray}
\boldsymbol{v}(\boldsymbol{x}) = \frac{1}{Z(\boldsymbol{x})} A(\boldsymbol{x}) \boldsymbol{V}(\boldsymbol{X})
\end{eqnarray}
where the matrix $A(\boldsymbol{x})$ is given by:
\begin{eqnarray}
A(\boldsymbol{x}) = \begin{bmatrix}
       f & 0 & -x           \\[0.3em]
       0 & f & -y
     \end{bmatrix}.\nonumber
\end{eqnarray}

In the continuous formulation, the velocity of the point relative to the camera
$\boldsymbol{V}(\boldsymbol{X})$ arises from a combination of translational and
rotational motions,
\[
\boldsymbol{V}{(\boldsymbol{X})} = \boldsymbol{\tau}+\boldsymbol{X} \times \boldsymbol{\omega}
\]
where $\boldsymbol{\omega} = (\omega_x, \omega_y, \omega_z)^\top \in
\mathbb{R}^3$ is unit length axis representation of rotational velocity
of the camera and $\boldsymbol{\tau} = (\tau_x, \tau_y, \tau_z)^\top \in \mathbb{R}^3$ is 
the translation.  Denoting the inverse depth at image location
$\boldsymbol{x}$ by $\rho(x) = \frac{1}{Z(x)}$, we can see that the projected
motion vector $\boldsymbol{v}$ is a linear function of the camera motion
parameters:
\begin{eqnarray}
\boldsymbol{v}(\boldsymbol{x})  &=& \rho(\boldsymbol{x})  A(\boldsymbol{x})  \boldsymbol{\tau} + B(\boldsymbol{x})  \boldsymbol{\omega} \nonumber \\
&=& \begin{bmatrix}
       \rho(\boldsymbol{x})  A(\boldsymbol{x})  & B(\boldsymbol{x})          
     \end{bmatrix} 
	\begin{bmatrix}
       \boldsymbol{\tau} \\ \boldsymbol{\omega}
     \end{bmatrix}
     \nonumber\\
&=& Q(\boldsymbol{x}) \boldsymbol{T},     
     \nonumber
\end{eqnarray}
where the matrix $B$ includes the cross product
\begin{eqnarray}
B(\boldsymbol{x}) = \begin{bmatrix}
       -xy & f+x^2 & \,\, -y           \\[0.3em]
       -f-y^2 & xy & \, x
     \end{bmatrix}.
     \nonumber
\end{eqnarray}

To describe motion field for the whole image, we concatenate equations for all
$N$ pixel locations and write $\mathcal{U} = \mathcal{Q}\boldsymbol{T}$ where
\begin{eqnarray}
\mathcal{U} = \begin{bmatrix}
\boldsymbol{v}(\boldsymbol{x}_1) \\
\boldsymbol{v}(\boldsymbol{x}_2) \\
\vdots\\
\boldsymbol{v}(\boldsymbol{x}_N)
 \end{bmatrix}   \in \mathbb{R}^{2N \times 1 },
\mathcal{Q} = \begin{bmatrix}
\rho_1 A(\boldsymbol{x}_1) & B(\boldsymbol{x}_1) \\
\rho_2 A(\boldsymbol{x}_2) & B(\boldsymbol{x}_2) \\
\vdots & \vdots \\
\rho_N A(\boldsymbol{x}_N) & B(\boldsymbol{x}_N )
 \end{bmatrix}   \in \mathbb{R}^{2N \times 6 },
\boldsymbol{T}  \in \mathbb{R}^{6 \times 1 }. \nonumber
\end{eqnarray}
We assume the focal length is a fixed quantity and in the following write the
motion field as a function $\mathcal{U} = MF(\boldsymbol{\rho}, \boldsymbol{T})$
which is linear in both the inverse depths $\boldsymbol{\rho}$ and camera motion
parameters $\boldsymbol{T}$.

To infer the camera motion $\boldsymbol{T}$ given inverse depths
$\boldsymbol{\rho}$ and image velocities $\mathcal{U}$, we use a least-squares
estimate:
\begin{eqnarray}
\boldsymbol{t}^*, \boldsymbol{\omega}^* &=&  \arg \min_{\boldsymbol{t}, \boldsymbol{\omega}}\sum_i^N w(\boldsymbol{x_i}) || v(\boldsymbol{x}_i) - \frac{1}{Z(\boldsymbol{x}_i)}A(\boldsymbol{x}_i)t + B(\boldsymbol{x}_i)\omega  ||^2 \nonumber
\end{eqnarray}
where $w(\boldsymbol{x_i})$ is a weighting function that models the reliability
of each pixel velocity in estimating the camera motion. The solution to this
problem can be expressed in closed from using the pseudo inverse of matrix
$\mathcal{Q}$.  We denote the mapping from $\mathcal{U}$ to estimated camera
motion as $\boldsymbol{T} = MF^\dagger(\boldsymbol{\rho}, \mathcal{U},
\boldsymbol{w})$.

In our model we utilize $MF^\dagger(\boldsymbol{\rho}, \mathcal{U},
\boldsymbol{w})$ to estimate camera model and $MF(\boldsymbol{\rho},
\boldsymbol{T})$ to resynthesize the resulting motion field. Both functions are
differentiable with respect to their inputs (in fact linear in $\mathcal{U}$
and $\boldsymbol{T}$ respectively) making it straightforward and efficient to
incorporate them into a network that is trained end-to-end using gradient-based
methods.

\begin{figure}[t]
\begin{center}
\scriptsize
\begin{tabular}{c c @{\hspace{0.5em}} c}
\includegraphics[trim=260 100 260 100, clip, height=29mm]{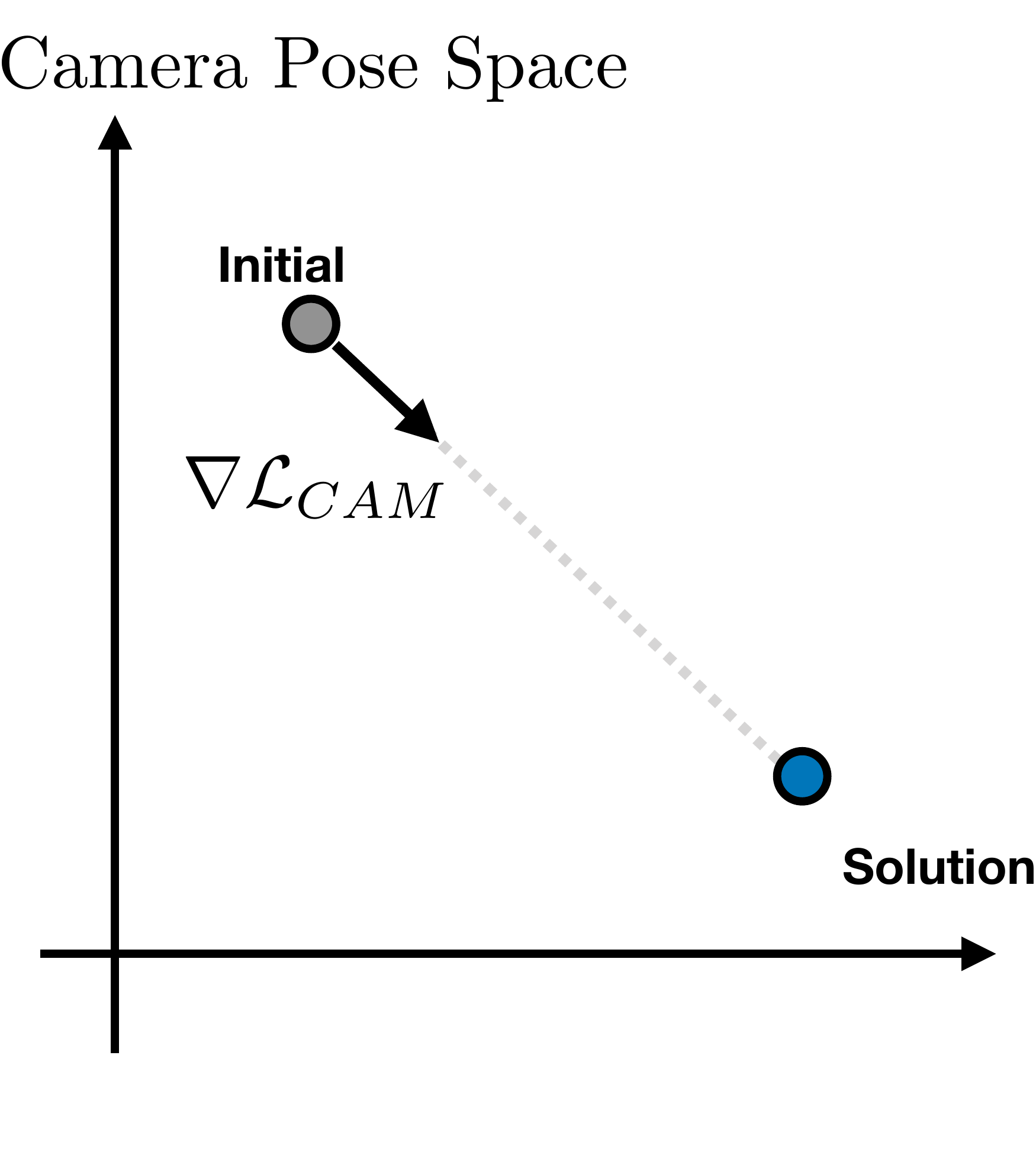}&
\includegraphics[trim=150 100 70 100, clip, height=29mm]{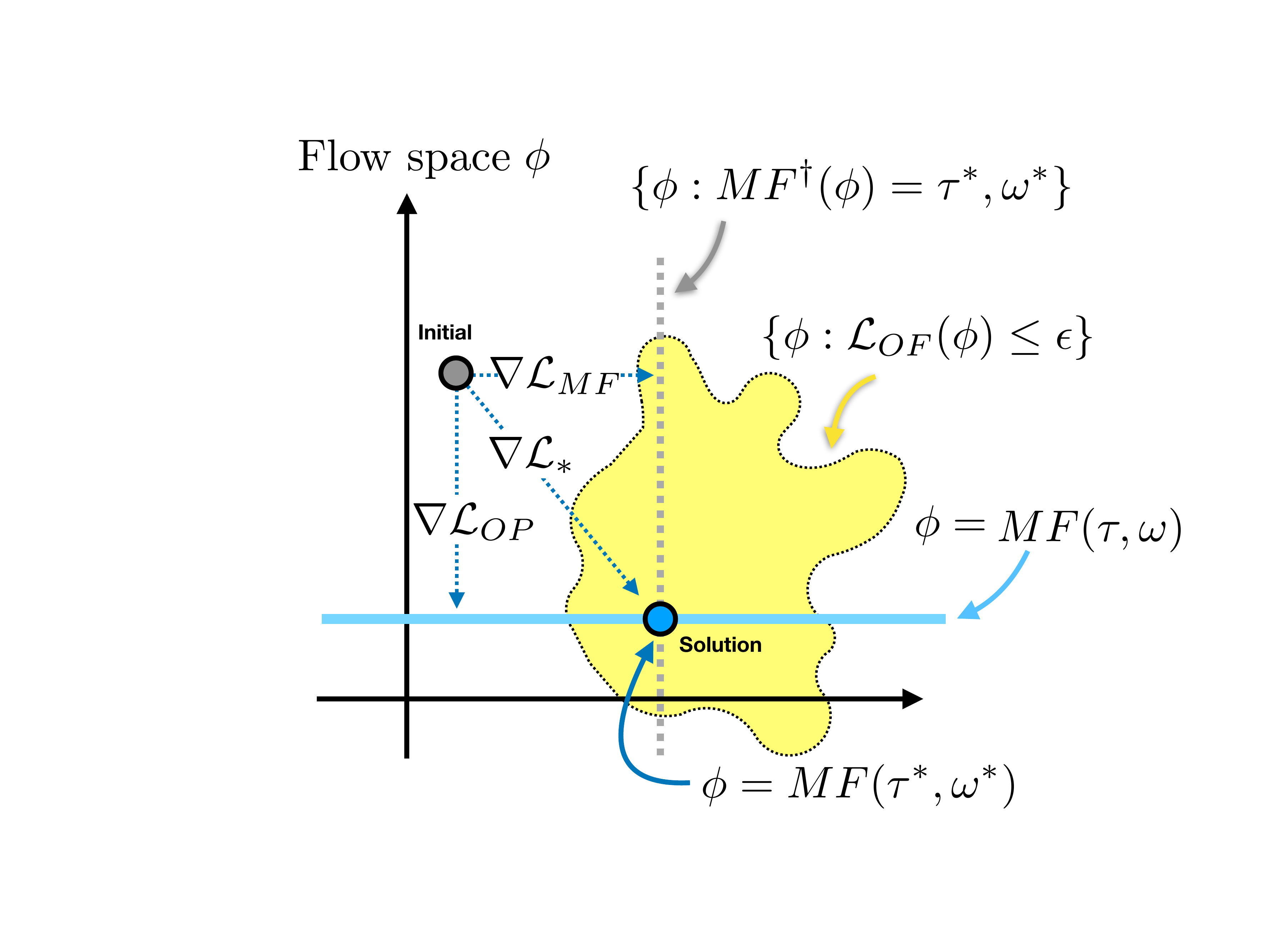}&
\includegraphics[trim=100 120 100 150, clip, height=29mm]{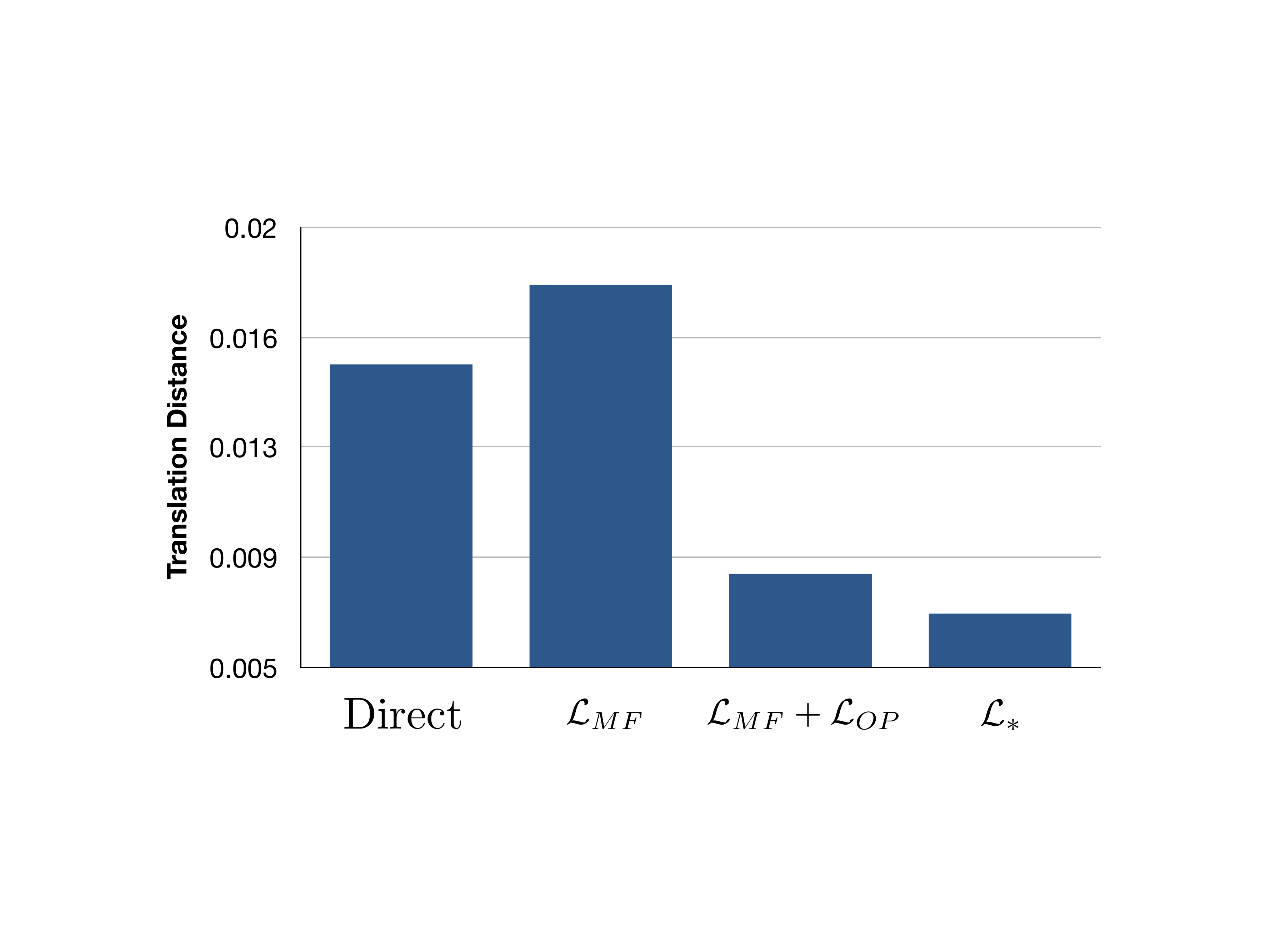}\\
(a) Predict pose directly & ~~~~(b) Predict pose via flow space & (c) losses comparison\\
\end{tabular}
\caption{Schematic interpretation of different loss functions. (a) Supervised
training of direct models utilize a loss defined on camera pose space.  (b) Our
approach defines losses on the space of pixel flows and considers losses that
measure the distance to the true motion field, the sub-space of possible
ego-motion fields (blue), and its orthogonal complement (gray dashed). The
model is also guided by photometric or scene-flow consistency between input
frames (yellow) (c) shows prediction error for supervised models trained with
different combinations of these losses and indicates that using losses defined
in flow-space outperforms direct prediction of camera motion. }
\label{fig:space}
\end{center}
\vspace{-0.1in}
\end{figure}

\def\tclip{100}
\def\bclip{20}
\def\hh{5.7em}

\begin{figure}[t]
\begin{center}
\scriptsize
\begin{tabular}{c@{\hspace{1.1em}}c@{\hspace{1.1em}}c@{\hspace{0.3em}}c}
\includegraphics[trim=640 {\tclip} 0 {\bclip}, clip, height={\hh}]{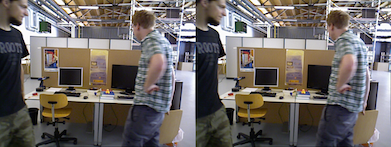}&
\includegraphics[trim=0 {\tclip} 0 {\bclip}, clip, height={\hh}]{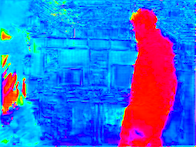}&
\includegraphics[trim=0 {\tclip} 0 {\bclip}, clip, height={\hh}]{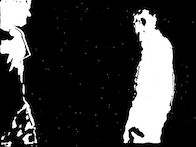}&
\includegraphics[trim=0 {\tclip} 0 {\bclip}, clip, height={\hh}]{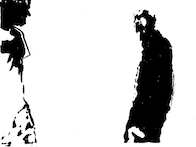}\\
(a) $I_t$ & (d) Optical Flow & (g) $\boldsymbol{Seg_{{dynamic}}}$ & (j) $\boldsymbol{Seg_{{static}}}$ \vspace{0.2em}\\
\includegraphics[trim=0 {\tclip} 640 {\bclip}, clip, height={\hh}]{./images/sample/00066002-frmae_389-inputs}&
\includegraphics[trim=0 {\tclip} 0 {\bclip}, clip, height={\hh}]{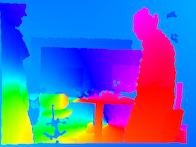}&
\includegraphics[trim=0 {\tclip} 0 {\bclip}, clip, height={\hh}]{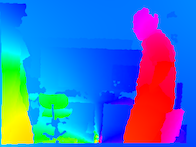}&
\includegraphics[trim=0 {\tclip} 0 {\bclip}, clip, height={\hh}]{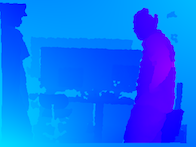}\\
(b) $I_{t+1}$ & (e) $\boldsymbol{MF_{global}}$ & (h) $\boldsymbol{MF_{dynamic}}$ & (k) $\boldsymbol{MF_{static}}$\vspace{0.2em}\\
\includegraphics[trim=0 {\tclip} 0 {\bclip}, clip, height={\hh}]{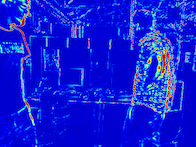}&
\includegraphics[trim=0 {\tclip} 0 {\bclip}, clip, height={\hh}]{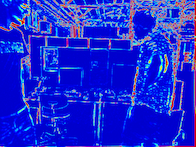}&
\includegraphics[trim=0 {\tclip} 0 {\bclip}, clip, height={\hh}]{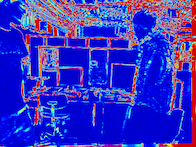}&
\includegraphics[trim=0 {\tclip} 0 {\bclip}, clip, height={\hh}]{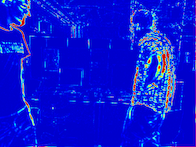}\\
(c) $|I_t - I_{t+\delta}|_1$ & (f) $|I_t - I_t^{\mathcal{W}_{global}}|_1$ & (i) $|I_t - I_t^{\mathcal{W}_{dynamic}}|_1$ & (l) $|I_t - I_t^{\mathcal{W}_{static}}|_1$\vspace{-0.2em}\\
\end{tabular}
\caption{A sample result on a dynamic sequence from TUM~\cite{sturm2012iros}.
From an input frame pair (a) and (b), $\boldsymbol{Net}_{OF}$ predicts optical
flow (d). Both camera and object motion are visible in the frame difference (c).
A single motion field (e) is dominated by large object motions and yields
poor warping error (f), particularly on the background.
Our model includes a segmentation network $\boldsymbol{Net}_{seg}$ that 
divides the image into dynamic and static masks (g,j) and fits corresponding
motion fields (h,k).  These provide better warping error on the objects (i)
and background (l) respectively.}
\label{fig:est_sample}
\end{center}
\vspace{-2.0em}
\end{figure}

\subsection{Projecting optical flow onto ego-motion}
\label{sec:motion_field_regen}

Given the true motion field $\mathcal{U}$, it is straight forward to estimate
the the true camera motion $\boldsymbol{T}^*$. In practice, the motion must be
estimated from image data which is often ambiguous (e.g., due to lack of
texture) and noisy. Typically there is a large set of image flows that are
photometrically consistent from which we must select the true motion field. Our
architecture utilizes a CNN to generate an initial flow estimate from image
data, then uses $MF^\dagger(\boldsymbol{\rho}, \mathcal{U}, \boldsymbol{w})$ to
fit a camera motion and finally reconstructs the image motion field
corresponding to the camera motion.  The composition of $MF^\dagger$ and $MF$
can be seen as a linear projection of the initial flow estimate into the space
of continuous motion fields.

A key tenet of our approach is that it is a better match to the capabilities
of a CNN architecture to predict the ego motion field in the image domain (and
subsequently map to camera motion) rather than attempting to directly predict
in the camera pose space. In particular, this allows for richer loss functions
that guide the training of the network.  We illustrate these idea schematically
for the case of supervised learning in~\figref{fig:space}.  Panel (a) depicts
the direct approach in terms of a loss function whose gradient pulls the
predicted pose towards the true pose.  

We display the relationship between optical flow, motion field and camera pose
in \figref{fig:space}(b). Among all possible image flows $\phi$, we indicate in yellow the set
which are photometrically valid (i.e., have a zero warping loss
$\mathcal{L}_{OF} \leq \epsilon$). The blue line indicates the 6-dimensional
subspace consisting of those motion fields that can be generated by all
possible camera velocities (conditioned on scene depth). Introducing a loss on
the camera pose (either directly on the prediction ${\tau,\omega}$, or on the
resynthesized motion field $MF(\tau,\omega)$ serves to pull the flow prediction
towards the orthogonal complement of this space (i.e., the set $\{\phi:
MF^\dagger(\phi) = {\tau^*,\omega^*}\}$ denoted by the gray vertical line).

Our approach allows the consideration of two other loss functions that can
provide additional guidance.  When supervision is available, we can introduce a
loss which directly measures the distance between the predicted flow and the
true motion field ($\mathcal{L}_{*}$ in the figure). In the self-supervised
setting, we can approximate this with the photometric warping loss
$\mathcal{L}_{OF}$.  Additionally, in either supervised or unsupervised
settings, we can include an orthogonal projection loss $\mathcal{L}_{OP}$,
which encourages the model to predict flows which are close to the space of
motion fields.  In section \label{sec:warping_loss}, we describe how these
losses are computed and adapted to the unsupervised setting.

While all of these losses are minimized in a perfect model,
\figref{fig:space}(c) shows that this choice of loss during training as a
substantial practical effect. In the supervised setting, optimizing the direct
loss in the camera pose space (using generic fully connected layers), or in the
flow space (using our least-squares fitting) results in similar prediction
errors. However, adding the projection loss or directly minimizing the distance
to the true motion field yields substantially better predictions (i.e., halving
average camera translation error).

\subsection{Static and Dynamic Motion Layers} 
So far, our description has assumed a camera moving through a single rigid
scene.  A standard approach to modeling non-rigid scenes (e.g., due to relative
motion of multiple dynamic objects in addition to ego-motion) is to split the
scene into a number of layers where each layer has a separate motion model
\cite{Weiss1997CVPR}.  For example, Zhou \emph{et al.} use a binary 
``explainable mask''~\cite{Zhou2017cvpr} to exclude outlying motions, and
Vijayanarasimhan \emph{et al.} segment images into K regions based on
motion~\cite{Vijayanarasimhan17Corr}. However, in the later-case, there is no
distinction between object motion and ego motion making it inappropriate for
odometry.

We use a similar strategy in order to separate motion into two layers
corresponding to static background and dynamic objects (outliers). We adopt
a u-net-like segmentation network~\cite{Ronneberger2015miccai} to predict
this separation which then defines the weights used for camera motion 
estimation using pseudo inverse function $MF^\dagger(\cdot)$ described in
\secref{sec:continuous_formulation}.

Consider a scene divided into $K$ regions corresponding to moving objects and
rigid background. Let $Seg_i(x) \in \{0,1\}$ denote a mask that indicates the
image support of region $i$ and $\mathcal{U}^i$ denote the corresponding rigid
motion field for that object considered in isolation. The composite motion
field for the whole image $\mathcal{U}$ can be written as:
\[
\mathcal{U} = \sum_i^K Seg_i \cdot \mathcal{U}^{i}\nonumber,
\]
In the odometry setting, we are only interested in the motion of the camera
relative to static background. We thus collect any dynamic objects into a
single motion field and consider a single binary mask:
\[
\mathcal{U}(\boldsymbol{x}_i) \approx Seg_{s}(\boldsymbol{x}_i)\mathcal{U}^{s}+Seg_{d}(\boldsymbol{x}_i)\mathcal{U}^{d} \nonumber.
\]
In our training with this segmentation network, we use the approximated motion
field $\mathcal{U}$ for the photometric warping loss described below.
For simplicity, we refer our single layer model as $\text{CeMNet}^1$ and dual
layer model as $\text{CeMNet}^2$

In ~\figref{fig:est_sample}, we illustrate intermediate results demonstrating
how the two layered model can better estimate camera motion in the presence of
dynamic objects. Since the single layer model cannot distinguish background and
foreground, the quality of predicted camera pose is bad. Excluding the dynamic
scene components from the camera motion estimation provides substantially better
pose estimation as seen in panels (i) and (l) which show less photometric 
warping error on the scene background relative to the single layer model
shown in (f).

\noindent\textbf{Hard assignment to layers:}
Previous work such as ~\cite{Vijayanarasimhan17Corr} uses a soft probabilistic
prediction of layer membership (i.e., using a softmax function to generate
layer weights).  However, such an approach introduces degeneracy since it can
utilize weighted combinations of two motions to match the flow (e.g., even in a
completely rigid scene). We find that using hard assignment of motions to
layers yields superior camera motion estimates. We utilize the ``Gumbel
sampling trick'' described in~\cite{Veit2017Corr} to implement hard assignment
while still allowing differentiable end-to-end training of both the flow and
segment networks.

\section{Training Losses}

\subsection{Losses for Self-supervision}
\label{sec:warping_loss}
As described in~\secref{sec:motion_field_regen}, there are several different
losses which can be applied to predicted flows. Here we adapt them to the
self-supervised setting.  The basic building block is to check if a 
predicted flow is photometrically consistent with the input image pairs.

For a given optical flow $\mathcal{U}^{OF}$ and source image $I_{t+\delta}$
we can synthesize warped image $I^{\mathcal{W}_{OF}}_{t}$ and check if it
matches $I_t$. As described in~\cite{Max2015NIPS}, this type of spatial 
transformation can be carried out in a differentiable framework using 
bilinear interpolation:
\[
I_t^{\mathcal{W}^{OF}}(x_i) = \sum_{i\in\{t,b\},j\in\{l,r\}} w^{ij}I_{t+\delta}(x_i+\mathcal{U}^{OF}(x_i)),\nonumber
\]
where $w^{ij}$ denotes the bilinear weighting of the four sample points. For
simplicity, we write $I_t^{\mathcal{W}^{OF}}(x_i) = \mathcal{W}(I_{t+\delta},
\mathcal{U}^{OF})$ to denote the warping of $I_{t+\delta}$ using flow
$\mathcal{U}^{OF}$. We then define the self-supervised flow loss using the
photometric error over all pixels: 
\[
\mathcal{L}_{OF} = \sum_{i}^N || I_t(x_i) - I^{\mathcal{W}^{OF}}_{t}(x_i)  ||_1\nonumber
\]
This loss serves as an approximation of $\mathcal{L}_*$ when the predictions are
far from the true motion field.

We can similarly apply warping loss to the reconstructed motion
field rather than the initial prediction.  If the motion field we found is correct, 
then again, the warped image should be matched with the target image. We can
build motion field loss by using motion-field warped image
$I_t^{\mathcal{W}^{MF}} = \mathcal{W}(I_{t+\delta}, \mathcal{U}^{MF})$ as:
\[
\mathcal{L}_{MF} = \sum_{i}^N P_t(x_i) || I_t(x_i) - I^{\mathcal{W}^{MF}}_{t}(x_i)  ||_1\nonumber
\]
where the mask $P_t(x_i)$ is 1 when the depth at $x_i$ is valid, 0 otherwise.
This is necessary when using a depth sensor which doesn't provide depths at
every image location.  This loss acts as a proxy for minimizing the camera
motion estimation error by lifting the prediction back to the flow space.  When
we predict camera motion for static scene, we use global motion field, and for
the dynamic scene, we use composite motion field $\mathcal{U}$.

\begin{figure}[t]
\begin{center}
\scriptsize
\def \w {8.2em}
\begin{tabular}{c@{\hspace{0.1em}}c@{\hspace{0.5em}}c@{\hspace{0.1em}}c@{\hspace{0.5em}}c}
\includegraphics[trim=0 0 640 0, clip, width=\w]{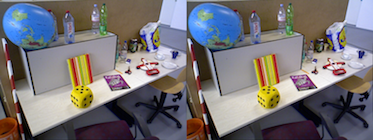}&
\includegraphics[trim=640 0 0 0, clip, width=\w]{./images/results_vis/00050001-frmae_179-inputs.png}&
\includegraphics[width=\w]{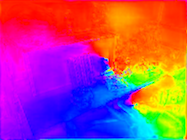}&
\includegraphics[width=\w]{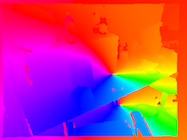}&
\includegraphics[width=\w]{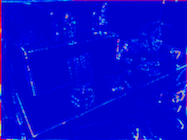}\\
\includegraphics[trim=0 0 640 0, clip, width=\w]{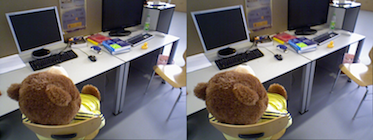}&
\includegraphics[trim=640 0 0 0, clip, width=\w]{./images/results_vis/00050001-frmae_1259-inputs.png}&
\includegraphics[width=\w]{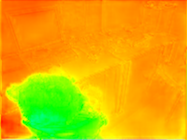}&
\includegraphics[width=\w]{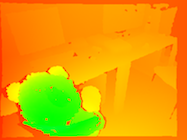}&
\includegraphics[width=\w]{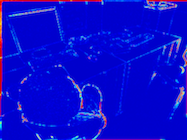}\vspace{0.5em}\\
\includegraphics[trim=0 0 1242 0, clip, width=\w]{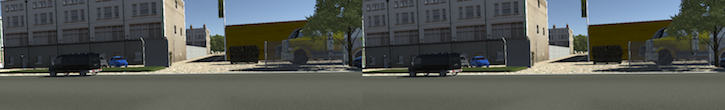}&
\includegraphics[trim=1242 0 0 0, clip, width=\w]{./images/results_vis/00120000-frmae_179-inputs.png}&
\includegraphics[width=\w]{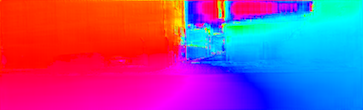}&
\includegraphics[width=\w]{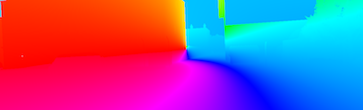}&
\includegraphics[width=\w]{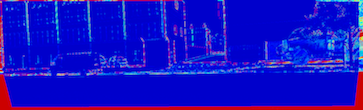}\\
\includegraphics[trim=0 0 1242 0, clip, width=\w]{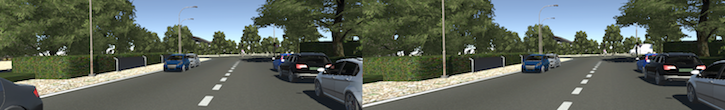}&
\includegraphics[trim=1242 0 0 0, clip, width=\w]{./images/results_vis/00120000-frmae_329-inputs.png}&
\includegraphics[width=\w]{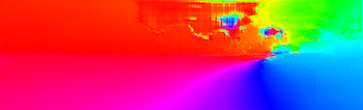}&
\includegraphics[width=\w]{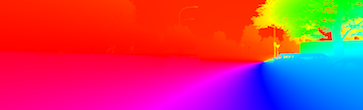}&
\includegraphics[width=\w]{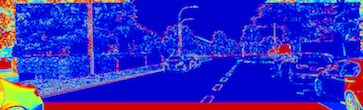}\\

(a) $I_t$ & (b) $I_{t+\delta}$ & (c) Optical flow & (d) Motion field & (e) $|I_t - I^{\mathcal{W}_{MF}}_t|_1$\vspace{-0.4em}
\end{tabular}
\caption{Visualizations of our single layered model.  Top three rows come from
TUM~\cite{sturm2012iros} dataset and bottom three come from Virtual
KITTI~\cite{Gaidon2016CVPR}. From the input images (a) and (b), the predicted
flow, and recovered motion field are displayed in (c) and (d) respectively.
Since motion field is derived from camera pose estimate, the error between
$I_t$ and motion field based warped image $I^{\mathcal{W}_{MF}}_t$ reflects the
accuracy of predicted camera motion.  If the predicted camera pose and depth is
ideal, then the error in (e) should be zero.
}
\label{fig:mf_warp_results}
\end{center}
\vspace{-1.0em}
\end{figure}

\begin{figure}[t]
\begin{center}
\begin{tabular}{c c}
\includegraphics[trim=30 150 30 100, clip, scale=0.16]{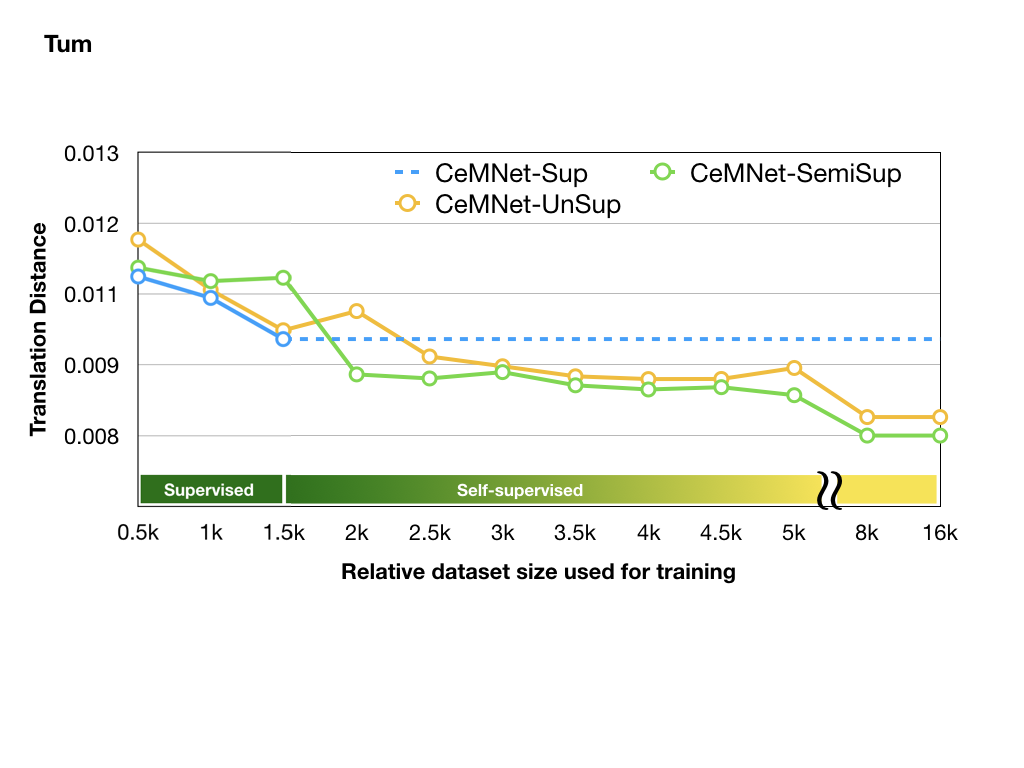}&
\includegraphics[trim=30 150 30 100, clip, scale=0.16]{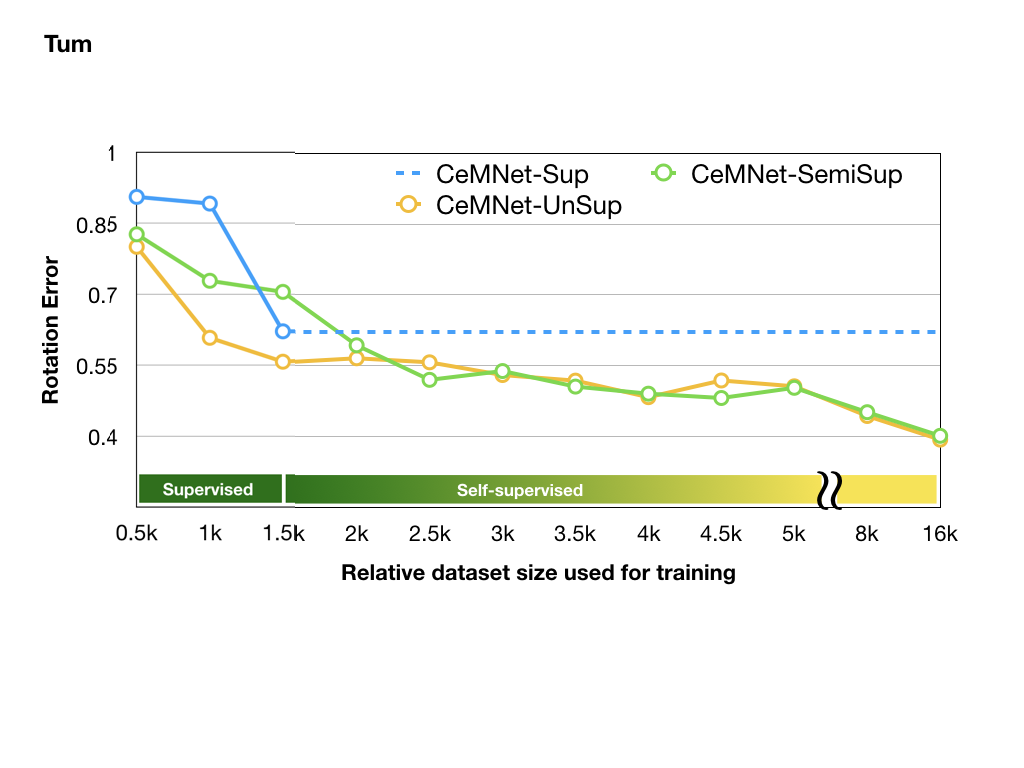}\vspace{-0.5em}\\
\includegraphics[trim=30 150 30 100, clip, scale=0.16]{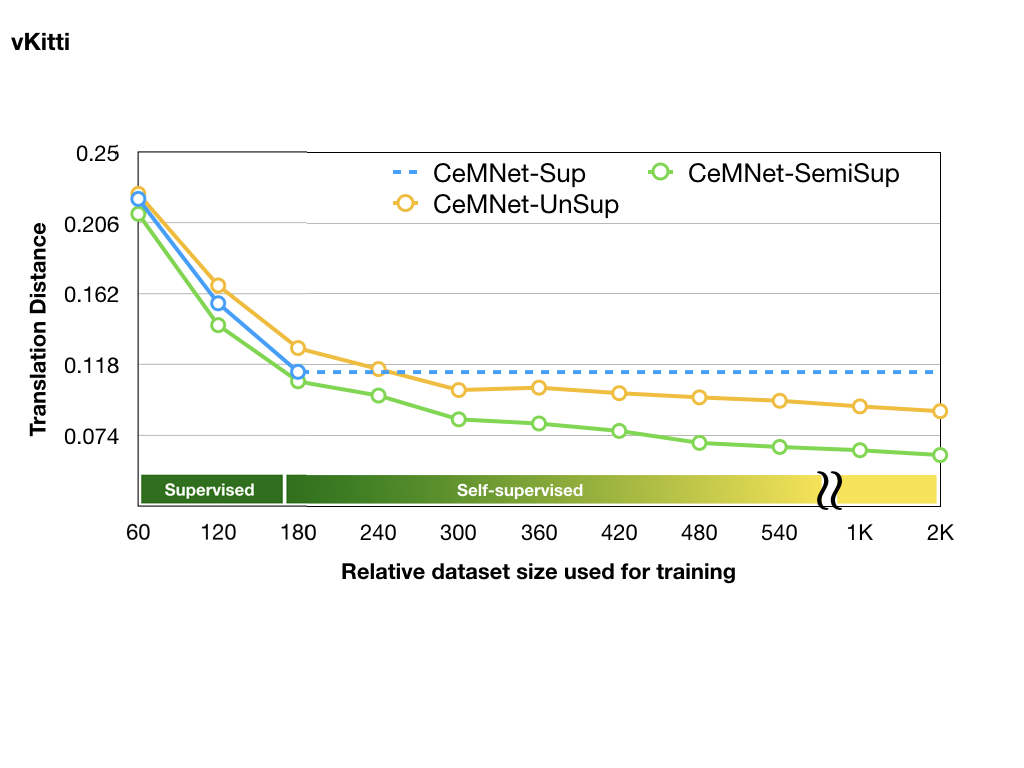}&
\includegraphics[trim=30 150 30 100, clip, scale=0.16]{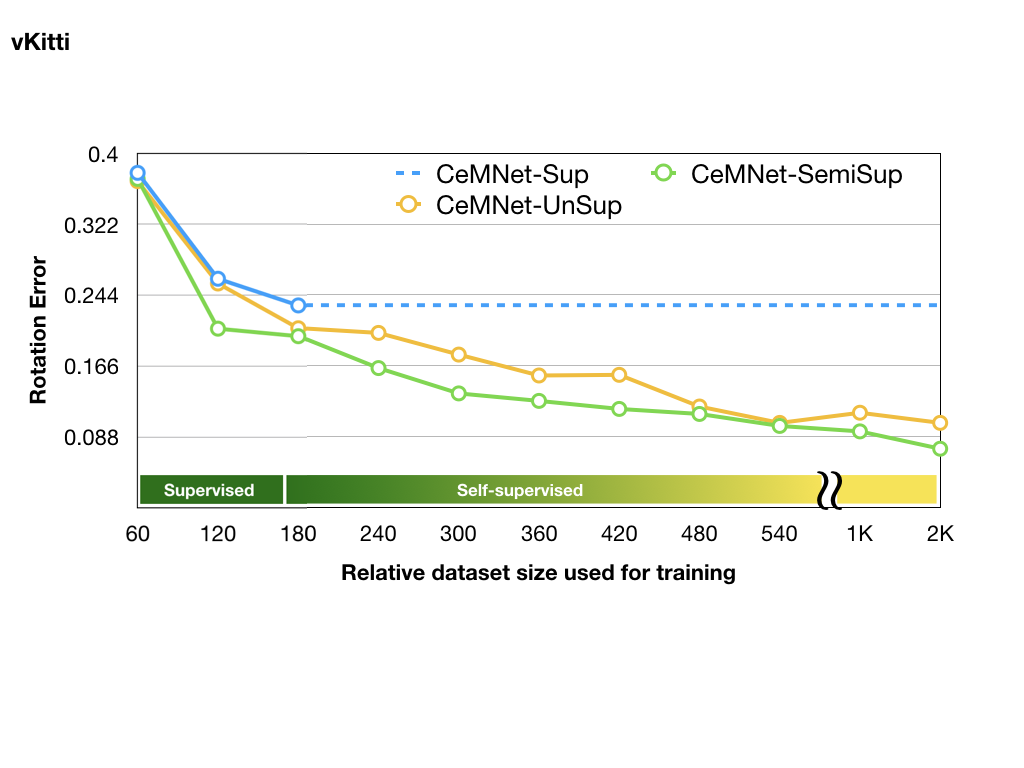}\vspace{-0.5em}\\
\scriptsize{(a) Translation Error} & \scriptsize{(b) Rotation Error}
\end{tabular}
\caption{Camera motion error on held-out test data as a function of
training set size for TUM (top) and Virtual KITTI (Bottom) RGBD datasets.
The blue line denotes training a supervised model that can't exploit unlabeled
data. The introduction of self-supervised warping losses yields much better
performance when either using only unsupervised training (yellow line) or
semi-supervised training (green).}
\label{fig:graph_dataset}
\end{center}
\vspace{-1.0em}
\end{figure}

Finally, we can utilize the orthogonal projection loss to minimize the distance
between predicted optical flow and its projection onto the space of motion fields
via:
\[
\mathcal{L}_{OP} = \sum_{i}^N || \mathcal{U}^{OF} - \mathcal{U}^{MF}  ||_1\nonumber
\]
By combining three above losses, we can define the final self-supervised loss 
function
\[
\mathcal{L}_{Final} = \lambda_{OF}\mathcal{L}_{OF} + \lambda_{MF}\mathcal{L}_{MF} +  \lambda_{OP}\mathcal{L}_{OP}, \nonumber
\]
where $\lambda_{OF}$, $\lambda_{MF}$ and $\lambda_{OP}$ weigh relative
importance (we use 1, 0.1 and 0.1 respectively in our experiments).

\subsection{Semi-supervision for symmetry breaking}
In our segmentation network, we have two layers corresponding to static and
dynamic parts.  However, in the unsupervised setting, the loss is symmetric
with respect to which is selected as background.  This symmetry problem can
interfere with training of the model and affect final performance. To break
this symmetry, we found it most effective to utilize a small amount of 
supervised data where camera motion is known.  For the supervised data we
use an additional loss term on the camera parameters.

\subsection{Camera Supervision from axis-angle representation}
Our network predicts camera motion in an axis-angle
representation that includes translation part $\boldsymbol{t} \in \mathbb{R}^3$ and
rotation $\boldsymbol{w} \in \mathbb{R}^3$. For supervised loss, we treat
these two components separately in order to match the criteria typically 
used in benchmarking pose estimation performance.

We first convert the axis-angle representation to a rotation matrix using
quaternions and then combine with the translation velocity to yield a
transformation matrix $\boldsymbol{R} \in \mathbb{R}^{4 \times 4}$.
Following~\cite{sturm2012iros}, we compute the difference between our predicted
transformation and the ground truth $\boldsymbol{R}^{d} =
(\boldsymbol{R}^{p})^{-1}\boldsymbol{R}^{gt}$ and penalize the translation and
rotation components respectively by:
\begin{eqnarray}
\mathcal{L}_{trans} &=& ||R^d_t||_2 \nonumber\\
\mathcal{L}_{rot} &=& \arccos\bigg(\min\bigg(1, \max\bigg(-1, \frac{Tr(R^d_r)-1}{2}\bigg)\bigg)\bigg)\nonumber
\end{eqnarray}

\begin{table}[t]
\begin{minipage}[t]{0.65\linewidth}
\resizebox{\linewidth}{!}{ 
\begin{tabular}{l@{\hspace{1em}} c @{\hspace{1em}} c @{\hspace{1em}} c@{\hspace{1em}} c@{\hspace{3em}} c}
\hline
\hline
\multirow{2}{*}{\scriptsize{Seq.}} & \scriptsize{DVO-SLAM} & \scriptsize{Kintinuous} & \scriptsize{ElasticFusion} & \scriptsize{ORB2}  & \scriptsize{CeMNet\tiny{(RGBD)}}\\
& \scriptsize{\cite{kerl13iros}} & \scriptsize{\cite{Whelan2012RSS}} & \scriptsize{\cite{WhelanRSS15}} & \scriptsize{\cite{MurArtal2017tor}}  & \\
\hline
\hline
fr1/desk & 0.021 & 0.037 &  0.020 &  0.016 &   \textbf{0.0089} \\
fr1/desk2 & 0.046 & 0.071 & 0.048 &  0.022 &     \textbf{0.0129} \\
fr1/room & 0.043  &  0.075& 0.068 &  0.047 &    \textbf{0.0071} \\
fr2/xyz & 0.018 & 0.029 & 0.011 &  0.004 &  \textbf{0.0009} \\
fr1/office & 0.035 & 0.030 & 0.017 &  0.010 &   \textbf{0.0041} \\
fr1/nst & 0.018 & 0.031 & 0.016 &  0.019&  \textbf{0.0117} \\
fr1/360 & 0.092 & - & - & -&  \textbf{0.0088} \\
fr1/plant & 0.025 &- & - & -& \textbf{0.0061}  \\
fr1/teddy & 0.043 &- & - & -&  \textbf{0.0139} \\
\hline
\hline
\end{tabular}}\vspace{1mm}
\end{minipage}
\hspace{1.0em}
\begin{minipage}[c]{0.3\linewidth}
\vspace{-0.5em}
\begin{center}

\caption{\small Relative translation error on TUM~\cite{sturm2012iros} static dataset.
Most of the methods in this table use RGBD frames camera for pose prediction.}
\label{tbl:compare_one_rgbd}
\end{center}
\end{minipage}
\vspace{-1.0em}
\end{table}

\begin{table}[t]
\begin{minipage}[t]{0.55\linewidth}
\resizebox{\linewidth}{!}{ 
\setlength{\tabcolsep}{.5em}
\begin{tabular}{l|cc|cc|cc}
\hline
\hline
\multirow{2}{*}{\scriptsize{Seq.}}\T & \multicolumn{2}{c|}{\scriptsize{TUM~\cite{sturm2012iros}}} &\multicolumn{2}{c|}{\scriptsize{SfM-Net~\cite{Vijayanarasimhan17Corr}} } & \multicolumn{2}{c}{\scriptsize{CeMNet\tiny{(RGB)}}}\\

& \scriptsize{Trans}\T& \scriptsize{Rot} & \scriptsize{Trans} & \scriptsize{Rot} & \scriptsize{Trans} & \scriptsize{Rot}\\
\cmidrule(r){1-1} \cmidrule(lr){2-3} \cmidrule(lr){4-5} \cmidrule(lr){6-7}
fr1/desk & \textbf{0.008} & 0.495 & 0.012 &  0.848  & 0.0113 & \textbf{0.6315}\\
fr1/desk2 & 0.099 & 0.61 & \textbf{0.012} & 0.974  & 0.0133 & \textbf{0.7548}\\
fr1/360 & 0.099 & 0.474 & \textbf{0.009} & 1.123 & 0.0091 & \textbf{0.5455}\\
fr1/plant & 0.016 & 1.053 & 0.011 & 0.796 & \textbf{0.0083} & \textbf{0.5487} \\
fr1/teddy & 0.020 & 1.14 & 0.0123 & 0.877 & \textbf{0.0113} & \textbf{0.6460}\\
\hline
\hline
\end{tabular}
}
\end{minipage}
\hspace{1.0em}
\begin{minipage}[c]{0.4\linewidth}
\vspace{1.5em}
\caption{\small Comparison to RGB SLAM odometry. To compare with methods that only use
RGB, we train our model using monocular depth prediction ~\cite{laina20163DV}
instead of input depth.
}
\label{tbl:compare_one_rgb}
\end{minipage}
\vspace{-1.0em}
\end{table}

\begin{table}[t]
\scriptsize
\begin{minipage}[t]{0.55\linewidth}
\resizebox{\linewidth}{!}{ 
\begin{tabular}{l c @{\hspace{0.5em}} c}
\hline
Method & Seq.09 & Seq.10\\
\hline
ORB-SLAM (full) & 0.014 $\pm$ 0.008 & 0.012 $\pm$ 0.011 \\
\hline
ORB-SLAM (short) &  0.064 $\pm$ 0.141 & 0.064 $\pm$ 0.130\\
Mean Odom &  0.032 $\pm$ 0.026 & 0.028 $\pm$ 0.023\\
Zhou \emph{et al.}~\cite{Zhou2017cvpr} & 0.021 $\pm$ 0.017 & 0.020 $\pm$ 0.015\\
Ours & 0.019 $\pm$ 0.014 & 0.018 $\pm$ 0.013\\
\hline
\end{tabular}}
\end{minipage}
\hspace{1.0em}
\begin{minipage}[c]{0.4\linewidth}
\vspace{1.0em}
\caption{\small Absolute Trajectory Error (ATE) comparison using KITTI dataset.
For this comparison, we average errors over 5-frame snippets.}
\label{tbl:compare_kitti}
\end{minipage}
\vspace{-3.0em}
\end{table}

\begin{table}[t]
\begin{minipage}[t]{0.55\linewidth}
\resizebox{\linewidth}{!}{ 
\begin{tabular}{c @{\hspace{.5em}} c @{\hspace{.5em}} c @{\hspace{.5em}} c @{\hspace{.5em}} c @{\hspace{1em}} c}
\hline
\hline
& \multicolumn{2}{c}{Training} \T&  Testing & &\\
\hline
 & \tiny{GT Depth} & \tiny{GT Cam} & \tiny{GT Depth} & Trans & Rot \\
\hline
Geometric~\cite{Jaegle2016icra} \T &- & - &   & 0.4579 & 0.3423\\
AIGN-SfM~\cite{Tung2017Corr} \T\B & \checkmark & \checkmark  &  & 0.1247  & 0.3333\\
CeMNet\tiny{(RGBD)} \T\B & \checkmark &  & \checkmark & \textbf{0.0878} & \textbf{0.0781} \\
CeMNet\tiny{(RGB)} \T &  & & & 0.0941 & 0.1079 \\
\hline
\hline 
\end{tabular}}
\end{minipage}
\hspace{1.0em}
\begin{minipage}[c]{0.4\linewidth}
\vspace{1.0em}
\caption{\small Relative pose error comparison using Virtual
KITTI~\cite{Gaidon2016CVPR}.
Both with (CeMNet{\scriptsize
(RGBD)}) and without (CeMNet{\scriptsize (RGB)}) depth inputs, our models
outperform previous methods.}
\label{tbl:compare_vkitti}
\end{minipage}
\vspace{-1.0em}
\end{table}

\section{Experimental Results}
\label{sec:results}
For the following experiments, we use the synthetic Virtual KITTI
dataset~\cite{Gaidon2016CVPR} depicting street scenes from a moving car, and
the TUM RGBD dataset~\cite{sturm2012iros} which has been used to benchmark a
variety of RGBD odometry algorithms.  To measure performance, we use relative
pose error protocol proposed in~\cite{sturm2012iros}. 

\vspace{0.1in}\noindent{\bf Self-supervised learning improves model performance:}
To show the benefits of self-supervision, we assume that only 10\% of each 
dataset has ground-truth available. We use 11 different sequences from the
TUM dataset as training, choose a random ordering of frame pairs over the
whole dataset and train models with increasingly large subsets of the data
and test on a separate held-out collection of frames.  This allows us to
evaluate the effect of growing the amount of supervised/unsupervised training
data in a consistent way across models.

In \figref{fig:graph_dataset}, we plot the relative translation/rotation errors
as a function of training data size.  The supervised version of the model
(CeM-Sup) can only be trained on the first 10\% of the dataset and makes
no use of the unsupervised data.  In this setting it outperforms the 
unsupervised model (CeM-Unsup).  However, as the amount of unsupervised 
training data continues to grow, CeM-Unsup eventually outperforms the
supervised model. For a clear comparison, the unsupervised losses are not used
in training (CeM-Sup). We also compare a model which uses both supervised
and unsupervised loss (CeM-SemiSup) which generally yields even better 
performance. We note that because the real world depth data in TUM is 
incomplete, limiting performance of the supervised model while the supervised
model shows expected decreasing errors on Virtual KITTI.
 
\vspace{0.1in}\noindent{\bf Motion field and warping:}
In~\secref{sec:warping_loss}, we describe how a predicted camera pose is used
to generate motion field and used in the warping loss.
In~\figref{fig:mf_warp_results}, we plot the per-pixel warping loss for several
inputs.  Left two (a-b) show the input RGB frames, (c) shows predicted optical
flow. (d) is regenerated motion field. (e) shows differences between the target
image and warped image. Note that blue color means lower differences between
those two images. 

\vspace{0.1in}\noindent{\bf Camera motion error comparison:}
To measure the quality of predicted camera pose, we compare our single layer
model (CeMNet) with previous RGBD SLAM methods on the TUM dataset
in~\tabref{tbl:compare_one_rgbd}.  CeMNet\small{(RGBD)} shows the best average
performance among tested methods in terms of relative translation error.
Several previous methods of interest,
including~\cite{Zhou2017cvpr,Vijayanarasimhan17Corr} do not utilize depth as
an input, instead predicting it directly from input images.

For fair comparison, we also test our model with predicted depth
(CeMNet\small{(RGB)}) using off-the-shelf the monocular depth prediction method
introduced by Iro \emph{et al.}~\cite{laina20163DV}.  This model was pretrained
using NYU Depth dataset V2~\cite{Silberman2012ECCV}. We rescale the predictions
by 0.9 to match the range of depths in TUM (presumably due to differences in
focal length) but otherwise leave the model fixed.
focal length for TUM.  As shown in~\tabref{tbl:compare_one_rgb}, our method
continues to outperform others in terms of rotation and shows comparable
translation errors.
As another comparison, we use KITTI~\cite{Geiger2012CVPR} dataset for
absolute trajectory error in~\tabref{tbl:compare_vkitti}. For training,
we use sequence from 00 to 08, and use 09 and 10 for each evaluation. 

Additionally, we show performance on the Virtual KITTI dataset
in~\tabref{tbl:compare_vkitti}. We specify how each method uses the available
ground truth depth and camera pose data available for train and test.
Using the true depth at test time results in strong performance from 
our model. For fair comparison, we also evaluate our model using the 
monocular depth prediction model of~\cite{Godard2017CVPR} pretrained with
KITTI~\cite{Geiger2012CVPR} dataset and converted from the predicted disparity
to depth\footnote{We use 0.54 as baseline distance and 725 for focal length}.
The results show better performance than previous self-supervised approaches
even without using ground-truth depth.

\begin{figure}[t]
\begin{center}
\scriptsize
\def \w {5.9em}
\begin{tabular}{c@{\hspace{0.2em}}c@{\hspace{0.5em}}c@{\hspace{0.2em}}c@{\hspace{0.5em}}
c@{\hspace{0.2em}}c@{\hspace{0.2em}}c}
\includegraphics[trim=0 0 640 0, clip, width=\w]{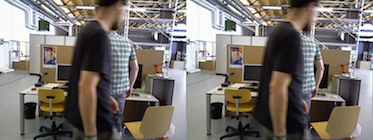}&
\includegraphics[width=\w]{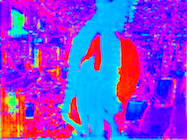}&
\includegraphics[width=\w]{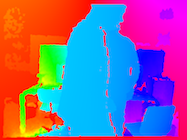}&
\includegraphics[width=\w]{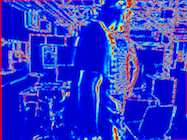}&
\includegraphics[width=\w]{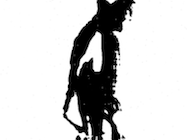}&
\includegraphics[width=\w]{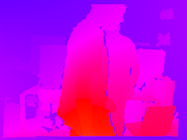}&
\includegraphics[width=\w]{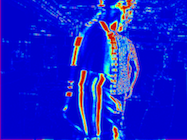}\\
\includegraphics[trim=0 0 640 0, clip, width=\w]{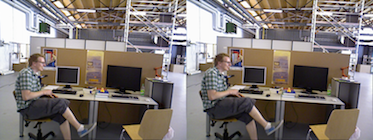}&
\includegraphics[width=\w]{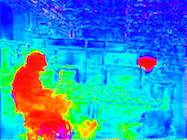}&
\includegraphics[width=\w]{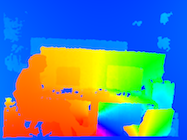}&
\includegraphics[width=\w]{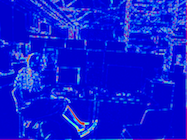}&
\includegraphics[width=\w]{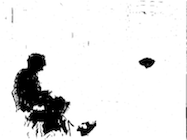}&
\includegraphics[width=\w]{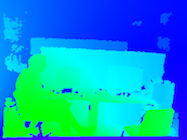}&
\includegraphics[width=\w]{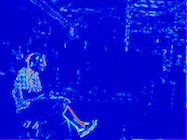}\\
\includegraphics[trim=0 0 640 0, clip, width=\w]{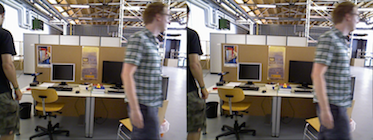}&
\includegraphics[width=\w]{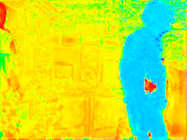}&
\includegraphics[width=\w]{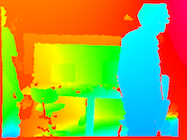}&
\includegraphics[width=\w]{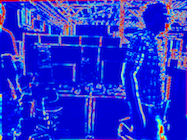}&
\includegraphics[width=\w]{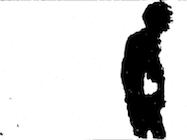}&
\includegraphics[width=\w]{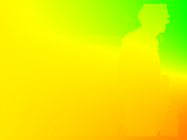}&
\includegraphics[width=\w]{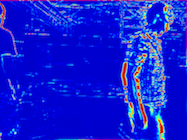}\\
(a) $I_t$ & (b) Optical & (c) $\boldsymbol{MF}$ & (c) $\boldsymbol{WE}$ & (d) $\boldsymbol{Seg}$ & (e) $\boldsymbol{MF}$ & (f) $\boldsymbol{WE}$\\
 & ~~~~flow & ~~~~(all) & ~~~~(all) & ~~~~~(static) & ~~~~~(static) & ~~~~~(static)\vspace{-0.5em}
\end{tabular}
\caption{Intermediate results of two layered model for dynamic scene camera
pose prediction. Without separating static and dynamic components, it is
difficult to get good camera motions (high error in (c)). However, as shown in
(f), it is possible to predict camera motion for background by fitting only
the static segment (d). }
\label{fig:seg_results}
\end{center}
\vspace{-0.5em}
\end{figure}

\begin{table}[t]
\begin{center}
\scriptsize
\setlength{\tabcolsep}{.7em}
\begin{tabular}{l|cc|cc|cc|cc}
\hline
\hline
\multirow{2}{*}{\scriptsize{Seq.}}\T & \multicolumn{2}{c}{\scriptsize{Baseline}}
 \T & \multicolumn{2}{c}{\scriptsize{$\text{CeMNet}^1$}} & \multicolumn{2}{c}
 {\scriptsize{$\text{CeMNet}^2$}} & \multicolumn{2}{c}
 {\scriptsize{$\text{CeMNet}^2$\tiny{(Semi)}}} \\
 & \scriptsize{Trans}\T & \scriptsize{Rot} & \scriptsize{Trans} & \scriptsize{Rot}  & \scriptsize{Trans} & \scriptsize{Rot} & \scriptsize{Trans} & \scriptsize{Rot}\\
\cmidrule(r){1-1} \cmidrule(lr){2-3} \cmidrule(lr){4-5} \cmidrule(lr){6-7} \cmidrule(lr){8-9}
fr3/sit\_static & 0.0134 & 0.5724 & 0.0025 & 0.1667 & 0.0016 & 0.1573 & \textbf{0.0010} & \textbf{0.1527}\\
fr3/sit\_xyz & 0.0179 & 0.7484 & 0.0070 & 0.2645 & 0.0068 & 0.2653 & \textbf{0.0064} & \textbf{0.2612} \\
fr3/sit\_halfsph & 0.0104 & 1.0135 & 0.0081 & \textbf{0.5272} & 0.0080 & 0.5820 & \textbf{0.0074} & 0.5552\\
fr3/walk\_static & 0.0149 & 0.5703 & 0.0103 & 0.2107 & 0.0030 & 0.1610 & \textbf{0.0019} & \textbf{0.1583}\\
fr3/walk\_xyz & 0.0174 & 0.7952 & 0.0128 & 0.3338 & 0.0079 & \textbf{0.2915} & \textbf{0.0078} & 0.2921\\
fr3/walk\_halfsph & 0.0166 & 0.9426 & 0.0147 & 0.4698 & 0.0107 & 0.4120 & \textbf{0.0102} & \textbf{0.3989}\\
\hline
\hline
\end{tabular}\vspace{1.5mm}
\caption{Relative pose error comparison using TUM dynamic
dataset~\cite{sturm2012iros}. Generally, the two layered
model shows better performance than single layered model.
Including a small amount of supervision ($\text{CeMNet}^2(Semi)$)
yields equivalent or better performance depends on dataset
by breaking the symmetry of the unsupervised loss.}
\label{tbl:compare_dynamic}
\end{center}
\vspace{-2.5em}
\end{table}

\vspace{0.1in} \noindent{\bf Static/Dynamic segmentation:}
In~\figref{fig:seg_results}, we visualize the results of breaking the input
into static and dynamic layers. From the RGB input pair at $I_t$ (a) and
$I_{t+\delta}$, predicted optical flow is shown in (b). While single layered
model generates motion field using the complete flow (c), two layered model
focuses on static region (d) and generates motion field by only using it (e).
The warping error from the total flow (c) is higher than (f) especially in
background region. 

We perform a quantitative comparison on the TUM dynamic dataset which includes
both object and camera motion.  The results results are shown
in~\tabref{tbl:compare_dynamic}.  While single layered models such as the
baseline direct prediction model and $\text{CeMNet}^1$ are sensitive to dynamic
objects, two layered model $\text{CeMNet}^2$ shows less pose error. However, as
noted previously, the unsupervised loss suffers from a symmetry as to which
layer correspond to ego-motion. We evaluate the use of a small amount of
supervised data (10\%) to break this symmetry in the segmentation prediction
network.  This yields the the lowest resulting motion errors across nearly all
test sequences.

\section{Conclusion}
In this paper, we have introduced a novel self-supervised approach for
ego-motion prediction that leverages a continuous formulation of camera motion.
This allows for linear projection of flows into the space of motion fields and
(differentiable) end-to-end training. Compared to direct prediction of camera
motion (both our own baseline implementation and previously reported
performance), this approach yields more accurate two-frame estimates of camera
motions for both RGBD and RGB odometry. Our model makes effective use of
self-supervised training, allowing it to make effective use of ``free''
unsupervised data. Finally, by utilizing a two-layer segmentation approach makes
the model further robust to the presence of dynamic objects in a scene which
otherwise interfere with accurate ego-motion estimation.

\noindent\textbf{Acknowledgements:} This project was supported by NSF grants
IIS-1618806, IIS-1253538 and a hardware donation from NVIDIA.

\bibliographystyle{splncs}
\bibliography{egbib}

\end{document}